\documentclass[a4paper]{VisionStyle}
\usepackage{epsfig}

\begin{document}

\title{New Constraints on Galaxy Evolution from the Optical Monitor on XMM-Newton}

\author{T.P.\,Sasseen\inst{1}, I.\,Eisenman,\inst{1}, K. Mason\inst{2}, \and 
the Optical Monitor Team}

\institute{ Dept. of Physics, University of California, Santa Barbara, CA 93106
\and Mullard Space Science Laboratory, University College London, 
Hombury St. Mary
Dorking, Surrey RH5 6NT, U.K. }

\maketitle 

\begin{abstract}

We use galaxies detected in a deep ultraviolet XMM-Newton Optical
Monitor image and a model that predicts UV galaxy counts based on
local counts and evolution parameters to constrain galaxy
evolution to Z=1.2.   The 17' square 2000 \AA\/ (UVW2 filter) image was
taken as part of the XMM-OM team's guaranteed time program.
We detect sources in this image to a flux limit of 2.7 $\times$ 10$^{-17}$
ergs cm$^{-2}$ s$^{-1}$ \AA$^{-1}$ (AB magnitude = 22).  
Since some of the sources may be
stars, we perform a number of checks, including shape, color and
implied distance to remove stars from the detected counts.  We find
galaxy number counts as a function of magnitude roughly in agreement
\cite*{tsasseen-F7:milliard92}, but again find these counts are 
in excess of evolution models.  The excess counts at faint magnitudes
may provide evidence for either a new population of galaxies emerging
around Z=0.7 or more dramatic evolution than some earlier
predictions.  The integrated light from the detected galaxies totals 32--36 
ph cm$^{-2}$ s$^{-1}$ \AA$^{-1}$\/ sr$^{-1}$, placing a firm lower limit on the 
integrated UV light from galaxies.

\end{abstract}

\section{Introduction}

Measurements of the integrated light from a galaxy at
2000 \AA\/ provides a fairly direct measure of the
instantaneous rate of star formation, since the massive
stars that provide most of this radiation are short-lived
compared with the age of the galaxy.  Knowledge of the star
formation rate also gives a measure
of the rate of heavy element production in a galaxy, or
in the universe when a large sample of galaxies are
measured (\cite{tsasseen-F7:madau96}).  The integrated light
from these galaxies contributes to the extragalactic
background light at ultraviolet wavelengths, whose main
sources are hot stars and active galactic nuclei.
Measurements of galaxy number counts in the ultraviolet
have been made by \cite*{tsasseen-F7:milliard92} using the
FOCA balloon-borne UV telescope, \cite*{tsasseen-F7:gardner00}
and \cite*{tsasseen-F7:hill97} using HST archival fields. 
These data have been interpreted with models that
predict number counts based on galaxy spectral 
energy distributions (SED's) and
luminosity functions, such as those of \cite*{tsasseen-F7:armand94}
and \cite*{tsasseen-F7:granato00}.  The total far-ultraviolet 
extragalactic background has been measured to be as high
as 500 ph cm$^{-2}$ s$^{-1}$ \AA$^{-1}$ and as low as 
30 ph cm$^{-2}$ s$^{-1}$ \AA$^{-1}$ (see review by
\cite{tsasseen-F7:bowyer91}).  Predictions for the number of galaxies
that might be detected in deep ultraviolet Optical Monitor (OM) images
are given by \cite*{tsasseen-F7:sasseen97}.  In this
paper, we detect galaxies in a deep UV image taken with
the Optical Monitor (OM) and use these galaxy number counts to
place constraints on galaxy luminosity evolution via a
a galaxy evolution model similar to that of 
\cite*{tsasseen-F7:armand94}.  We also find a lower limit to the galaxy
contribution to the extragalactic UV background.

\section{The Data}   

The OM 13 hr deep field (at J2000.0 13 34 37.00, +37
54 44.0) was observed for approximately 200 ks with
XMM-Newton around June 22, 2001.  Details
of the OM exposures used in this study are shown in 
Table~\ref{tab:tab1}. 

\begin{table}[h]
  \caption{OM images used in this study.} 
  \label{tab:tab1}
  \begin{center}
    \leavevmode
    \footnotesize
    \begin{tabular}[h]{lcl}
      \hline \\[-5pt]
      Filter & Central Wavelength &  Exposure Time \\[+5pt]
        &   (\AA) &  (ksec) \\[+5pt]
      \hline \\[-5pt]
      B & 4200 & 10 \\
      U & 3900 & 10 \\
      UVW1 & 3000 & 20 \\
      UVM2 & 2500 & 31.5 \\
      UVW2 & 2000 & 30 \\
      \hline \\
      \end{tabular}
  \end{center}
\end{table}

\begin{figure}[ht]
  \begin{center}
    \epsfig{file=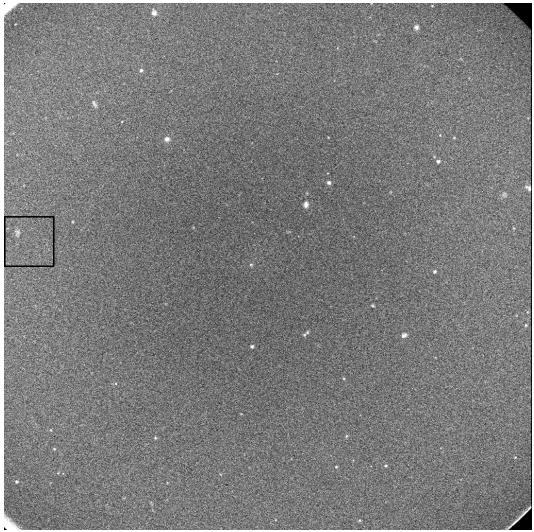, width=7cm}
  \end{center}
\caption{The 17' by 17' UVW2 (2000 A) image of
total exposure 31,450 s.
The five sigma detection limit determined with 5"
FWHM PSF is 3$\times 10^{15}$ ergs $cm^{-2} s^{-1}$.
We detected 122 sources above this limit.  Source with
further detail is indicated in box.}
\label{tsasseen-F7_fig1}
\end{figure}

\begin{figure}[ht]
  \begin{center}
    \epsfig{file=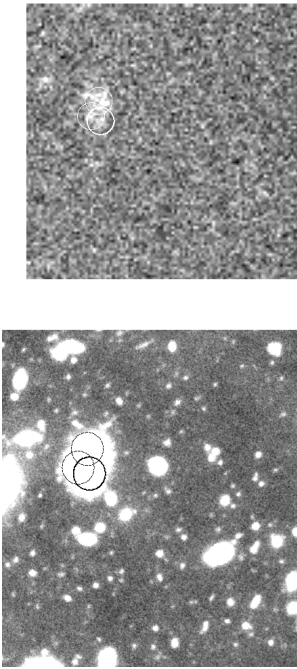, width=7cm}
  \end{center}
\caption{Detail of an example source region in UVW2 image (upper) and 
Subaru R-band image (lower)
with detection limit R$\sim$27. We examine multiple detections in one 
location in more detail, both in the direct images and in a surface plot
using the Subaru image.}
\label{tsasseen-F7_fig2}
\end{figure}

\begin{figure}[ht]
  \begin{center}
    \epsfig{file=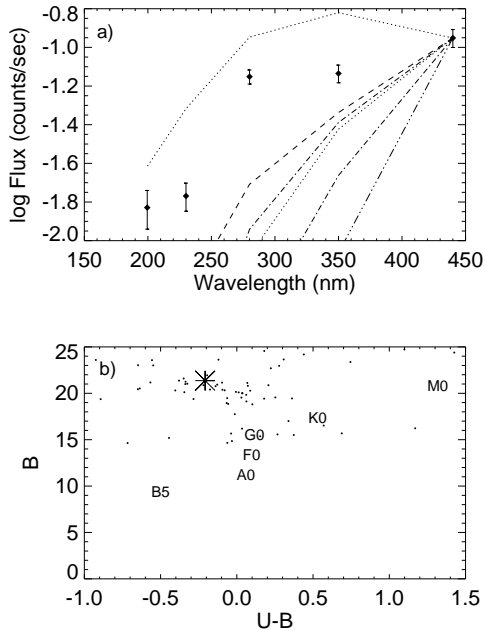, width=7cm}
  \end{center}
\caption{a) The spectral energy distribution of 
the example source measured with 
Optical Monitor filter bandpasses (points) plotted with template
stellar SED's for B0, A0, F0, G0, K0 and M0 dwarf stars (curves) for 
comparison.  The hot galaxy SED's is evident here and in most cases. 
b) A distance estimate based on U-B colors if the object
were a main-sequence star.  In the second plot, the
stellar type labels are shown at the position corresponding to 
where a main sequence star of that stellar type at a distance of 1 kpc 
would lie.  In most cases, these two figures allowed satisfactory 
discrimination between stars and galaxies.  We also referred to a 
surface plot of the source location extracted from the Subaru image.} 
\label{tsasseen-F7_fig3}
\end{figure}

Several exposures of typically 7 ks were 
brought to a common astrometric reference frame 
and coadded.  We
searched each image for sources using Sextractor and
made a catalog of the sources we found.  We concentrate
here on sources in the UVW2 image (Fig.~\ref{tsasseen-F7_fig1}) 
and use measurements in the other filters to differentiate between
stars, galaxies and QSO's.  We also use a deep R band
image (to R$\sim$27) of this field taken with the 8m Subaru 
telescope on Mauna Kea (Fig.~\ref{tsasseen-F7_fig2}) 
to check for source shape and possible confusion.  We 
perform two checks to discriminate stars from galaxies.
First, we compare the SED of each
UVW2 source (determined from OM photometry)
against stellar templates. Second, we compute an inferred
distance, as if the source were a main sequence star, 
from U-B color and B magnitude, as shown in Fig.~\ref{tsasseen-F7_fig3}.  
We find these checks form reliable stellar
discriminators for more than 90\% of the sources brighter
than AB=22.  \footnote{The AB magnitude system is defined by 
\newline
AB = $-2.5log(F_\nu) - 48.6$ where $F_\nu$ is given in 
ergs cm$^{-2}$ s$^{-1}$ Hz$^{-1}$ (\cite{tsasseen-F7:oke71}).}  
We also find a number of QSO's in the
field that show UV excess and
appear point-like in the OM and Subaru images.  We
categorize these separately in our galaxy number counts.  
Further work remains to completely discriminate any
remaining stellar content and the QSO populations.

We plot the detected galaxy counts as a function of
magnitude in Fig.~\ref{tsasseen-F7_fig4}. 
Our counts are in approximate agreement with that of
\cite{tsasseen-F7:milliard92} (also shown in Fig.~\ref{tsasseen-F7_fig4}) 
in the range of overlap, and we extend these counts to AB=22.

\section{The Models}

\begin{figure}[ht]
  \begin{center}
    \epsfig{file=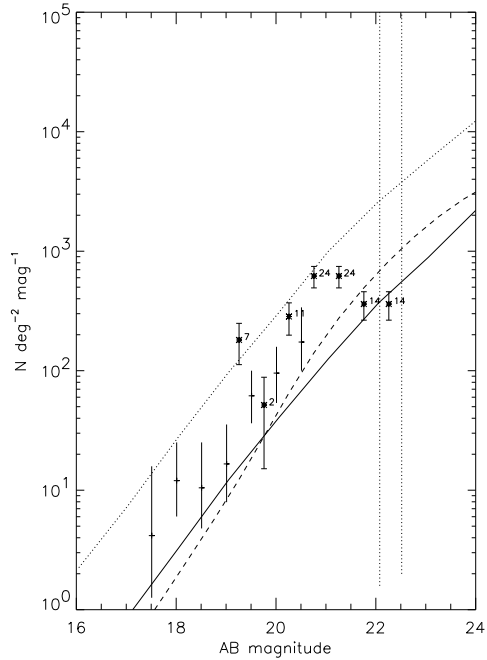, width=7cm}
  \end{center}
\caption{Our measured galaxy counts (starred points) after 
removing probable stellar sources.  The number indicates
the number of galaxies in that bin.  We also
show the data of Armand \& Milliard (1994) in a similar
bandpass, which are in rough agreement with 
galaxy counts in this study.  We
show our model results, as the dotted line.  We also show
the modeling results of Granato et al. (2000), where the dotted line 
is the model without dust included, and the solid line includes 
the effects of dust. The vertical dotted lines indicate the 3-$\sigma$
and 2-$\sigma$ detection levels in our UV image.} 
\label{tsasseen-F7_fig4}
\end{figure}

We have constructed a model is similar to that of \cite*{tsasseen-F7:armand94}
and use it to predict galaxy counts at 2000 \AA\/ as a function
of apparent magnitude. The model
uses a Schechter absolute luminosity distribution
function for 6 different galaxy types at redshifts between
zero and 1.2, along with K-corrections and a single
parameter luminosity evolution factor for each galaxy
type.  We have normalized the Schechter function using
observed counts at Bj=17, and set our evolution
parameters to agree with the modeled galactic evolution
of \cite*{tsasseen-F7:rocca88}, following
\cite*{tsasseen-F7:armand94}.  Our model implicitly includes the effects
of dust absorption and scattering because it is based on observed
UV SED's.  Like Armand \& Milliard,
our model predicts fewer galaxies in each magnitude
band than our measured number counts, as shown in
Figure~\ref{tsasseen-F7_fig4}.  
We also compare  the observed counts with
the model of \cite*{tsasseen-F7:granato00}, whose model
explicitly includes expected contributions to the observed galaxy
counts from starburst galaxies and dust.  Our model agrees well with
the Granato et al. model that includes dust, but our
observed counts are higher than both models that include dust.  

\section{Discussion} 

The summed the flux from non-stellar sources detected 
in the UVW2 image totals
32--36 ph cm$^{-2}$ s$^{-1}$ sr$^{-1}$ \AA$^{-1}$, 
with the higher limit including the contribution from
QSO's and active galaxies.
The integrated far-ultraviolet light from discrete galaxies
has been measured recently by \cite*{tsasseen-F7:gardner00} to be 
144--195 ph cm$^{-2}$ s$^{-1}$ sr$^{-1}$ \AA$^{-1}$, based on galaxies
detected in the range AB = 24 to 29.5 and a model to infer
the flux from brighter galaxies.  These authors claim there appears to 
be a break in the slope of the galaxy number counts that 
occurs around AB = 24, with substantial flattening of function
at fainter magnitudes.  Our measurements show an intriguing downturn
in galaxy counts at the faint end, which may indicate the 
start of the change in the slope of the number counts.  
There still remains some uncertainty in the number counts 
in the gap between our measurements
and those of \cite*{tsasseen-F7:gardner00}, 
which indicates the total integrated flux of galaxies 
is still uncertain.  

The discrepancy between the models shown in 
Fig.~\ref{tsasseen-F7_fig4}
and both our data and that of \cite{tsasseen-F7:milliard92} 
may indicate that we are missing some components in
our understanding of how galaxies evolve.  Some possible 
reasons for the descrepancy between their model and measurements
are given by \cite{tsasseen-F7:armand94}, including faster evolution of
the star formation rate or the possiblity that
there is a population of blue galaxies that is substantially
more numerous at z = 0.7 than they are today.  

\section{Future Work}

There are a number of effects we have not yet evaluated in detail
that may affect our measurement and conclusions.  These include
the effects of galaxy inclination, morphology and apertures on
our photometry, the effects of comparing measurements made
in slightly different bandpasses, 
and the detailed effects of dust absorption and possible 
evolution in galaxies (\cite{tsasseen-F7:vacca97}).  Our simple model assumes
a smooth evolution in star formation, but there is evidence that
star formation may be espisodic or occur in bursts, possibly 
because of merger activity, {\em e.g.} \cite*{tsasseen-F7:johnson98}.  
Galaxies change over time in many ways and our model predicts only
one facet of these changes, namely an evolving star formation rate.   
The full picture of galaxy evolution is certainly more complicated.  
It remains to explore further the connections between changes in the star
formation rate and changes in galaxy appearance and morphology,
metallicity, gas content, spectral energy output, and merger activity
that have been discussed at length by other researchers.  

\section{Conclusions}

1)  We have obtained galaxy counts at 2000 \AA\/ to a
magnitude of AB = 22 in deep images from the Optical
Monitor on XMM-Newton.  The long
OM exposure allows us to measure galaxy counts 1.5
magnitudes fainter than
\cite*{tsasseen-F7:armand94}, and we find similar counts
in range of overlap.
2)  Two evolutionary models underpredict
the observed galaxy counts, and may indicate that 
several process may be at work, including
episodic star formation, changes in the optical depth
within galaxies to
2000 \AA\/ radiation, or a new population of galaxies that is
less visible in the present epoch.
3)  The total integrated flux from the galaxies we detect to AB=22
is 32--36 ph cm$^{-2}$ s${-1}$ \AA$^{-1}$ sr$^{-1}$. 
This flux is a lower limit to the integrated 
extragalactic background light at 2000 \AA, and represents
about 20--25\% of the integrated, far-ultraviolet flux 
from galaxies inferred from the deep HST measurements of 
\cite{tsasseen-F7:gardner00}.

\begin{acknowledgements}

This research was supported by NASA grant NAG5-7714.
We would also like to thank the Optical Monitor team, 
and ESA for their successful program
to produce the first rate space observatory, XMM-Newton, 

\end{acknowledgements}

\end{document}